# Adaptive Drift-Diffusion Process to Learn Time Intervals

**Francois Rivest**[1]        **Yoshua Bengio**

Département d'informatique et de recherche opérationnelle

Université de Montréal

*francois.rivest@mail.mcgill.ca*      *bengioy@iro.umontreal.ca*

## Abstract

Animals learn the timing between consecutive events very easily. Their precision is usually proportional to the interval to time (Weber's law for timing). Most current timing models either require a central clock and unbounded accumulator or whole pre-defined populations of delay lines, decaying traces or oscillators to represent elapsing time. Current adaptive recurrent neural networks fail at learning to predict the timing of future events (the 'when') in a realistic manner. In this paper, we present a new model of interval timing, based on simple temporal integrators, derived from drift-diffusion models. We develop a simple geometric rule to learn 'when' instead of 'what'. We provide an analytical proof that the model can learn inter-event intervals in a number of trials independent of the interval size and that the temporal precision of the system is proportional to the timed interval. This new model uses no clock, no gradient, no unbounded accumulators, no delay lines, and has internal noise allowing generations of individual trials. Three interesting predictions are made.

## 1    Introduction

The ability to learn timing of events is crucial to survival. When a temporal relationship exists between two events, it seems that its timing is acquired almost as fast as the discovery of the relationship itself [1;2]. However, the neurobiological process underlying time interval learning remains unclear [3]. Three main possible adaptive neural representations of time intervals have been proposed [4]: (*i*) sequences of active neurons (so called *synfire* chains) [5], (*ii*) increasing numbers of active bistable neurons over time [6], or (*iii*) increasing neural activity within each individual neuron over time [7]. While synfire chains (*i*) are assumed to be more realistic for delays on the order of milliseconds [8], the other two options are more likely representations in the multi-seconds range. Models of temporal representation should also reproduce Weber's law for time, that is, that the timing precision of the model should be proportional to the timed interval. The growing population of active bistable neurons representation (*ii*) was recently shown to have this property [6] using both, abstract simple units, and a more realistic model of spiking neurons. But electrophysiological data suggesting that elapsing time could be represented through build-ups of activity within individual neurons (*iii*) seem more abundant [4;7;9-11]. While some realistic spiking neuron models of adaptive climbing activity exist [7], they are not used in the interval timing literature. In this paper, we present a simpler and more abstract neural model based on a drift-diffusion process of climbing neural activity. Such models are often used in decision making under noisy stimuli [12]. We extend them by developing a learning rule so that they can be used to learn time intervals rapidly. We show that such neural integrators could reproduce Weber's law for time, under the hypothesis that noise in synapses is related to the synaptic efficacies (weight).

---

[1] The corresponding author, Francois Rivest, is now at the Royal Military College of Canada, Department of Mathematics and Computer Science, francois.rivest@rmc.ca.

## 2    Methods

Consider a temporal conditioning situation: a situation in which once the animal is in the experimental chamber, rewards are delivered at a fixed rate. Let $x(t)$ be the presence (1) or absence (0) of stimulus X (the chamber) at time $t$. Let $y(t)$ be a sum of Dirac ($\delta(t)$) delta functions marking event Y onsets (reward delivery) at time $t_1, t_2, \ldots, t_n$ such that:

$$y(t) = \sum_{i=1}^{n} \delta(t_i) \tag{1}$$

where $t_i$ is the time at which the $i^{th}$ event occurs. Let $\varphi(t)$ be the output of a temporal integrator $\Phi$ predicting Y at time $t_i$ so that, starting from $\varphi(t_{i-1}+\varepsilon) = 0$, we want $\varphi(t_i) = 1$ and $0 \leq \varphi(t) \leq 1$, with infinitesimal small $\varepsilon > 0$. Let $w$ be the synaptic weight connecting stimulus X to the integrator $\Phi$.

### 2.1    System dynamics

The integrator dynamics are defined using the following equation:

$$d\varphi = 1_{\varphi \neq 1} xw d\tau - y\varphi d\tau \tag{2}$$

with the initial conditions $\varphi(0) = 0$ and $w(0) = \varepsilon$. The first term of (2) indicates that $\varphi$ integrates $xw$ over time until it reaches 1. The second term indicates that if event Y occurs, the temporal integrator is reset to 0. To simplify, we assume this reset is instantaneous. Assuming $w$ is constant, the integrator value $\varphi(t)$ at time $t$ is either equal to 1 or to the elapsed time in the presence of $x$ since last event (at time $t_i$) multiplied by the weight $w$. Thus, if $x = 1$ (animal in the chamber), $t_i < t < t_{i+1}$, and $w$ constant, then:

$$\int_{t_i}^{t} xw d\tau = (t - t_i)w \text{ and } \varphi(t) = \min\{(t - t_i)w, 1\} \tag{3}$$

For the remaining sections, we will assume $x = 1$ is always true to lighten notation.

### 2.2    Adaptation dynamics in the ideal case

We would like the temporal integrator to learn to predict the timing of the event Y by reaching its bound (1) at the moment Y occurs. If Y occurs at fixed intervals, then $w$ would encode exactly that interval:

$$\int_{t_i}^{t_{i+1}} xw d\tau = (t_{i+1} - t_i)w = 1 \Leftrightarrow w = \frac{1}{(t_{i+1} - t_i)} = \frac{1}{I_{i+1}} \tag{4}$$

where $I_i$ is the $i^{th}$ time interval (between event $i$-1 and $i$).

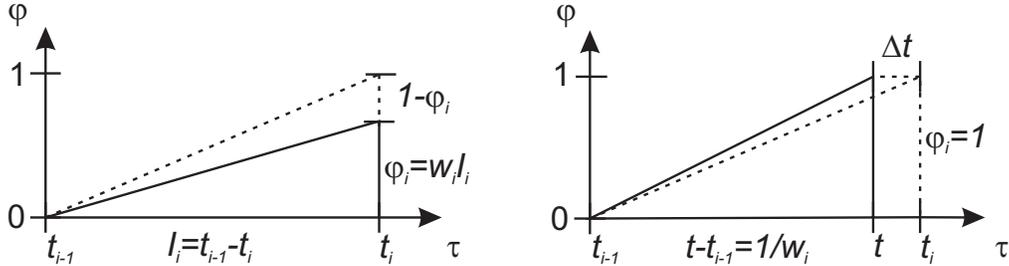

Figure 1: The geometry of the integrator when the event Y occurs before it has reached its bound (left panel) and when it reaches its bound before the event Y occurs (right panel). The trajectory is drawn in solid lines; dashed lines represent the desired trajectory.

Consider Y occuring sooner than expected, i.e., $t_i$ is such that $\varphi(t_i) < 1$ (Figure 1, left panel). Then $w$ can be corrected such that $\varphi$ reaches 1 at the next occurrence of Y (at time $t_{i+1}$), assuming Y occurs at constant time intervals, using the rule:

$$dw = yw\frac{(1-\varphi)}{\varphi}d\tau \tag{5}$$

This means that a discrete corrective jump in $w$ occurs at time $t_i$. It is easy to prove geometrically that this would correct the timing perfectly if the next event occurs at the same interval. The elapsed time since the last event $i$-1 can be estimated from $\varphi_i = \varphi(t_i)$ and $w_i = w(t_{i-1}+d\tau)$ (the value of $w$ during interval $I_i$, i.e. right after the update caused by the last event $i$-1), with $I_i = \varphi_i/w_i$. If the intervals $I_i$ are all of the same length, then a single trial is sufficient to correct the predicted time. On the next trial ($i$+1), $\varphi(\tau) = 1$ exactly when $\tau = t_{i+1}$:

$$\varphi_{i+1} = w_{i+1}I_{i+1} = w_{i+1}I_i = \left(w_i + w_i\frac{(1-\varphi_i)}{\varphi_i}\right)I_i = \varphi_i + \varphi_i\frac{(1-\varphi_i)}{\varphi_i} = 1 \tag{6}$$

The same correction rule cannot be applied if the integrator predicts the event Y too early. The boundedness of the integrator does not allow it to remember time beyond the expected time (Figure 1, right panel). On the other hand, when $\varphi$ reaches 1, the weight can be continuously decayed relative to the elapsed time since last event $i$ (i.e., relative to its own value) such that $w(t) = 1/(t-t_i)$ until $t = t_{i+1}$. The following decay maintains this property:

$$dw = -1_{\varphi=1}w^2 d\tau \tag{7}$$

Solving the differential equation and integrating over the period when $\varphi = 1$ gives:

$$dw = -w^2 d\tau \Rightarrow -\frac{dw}{w^2} = d\tau \Rightarrow \int_{w_i}^{w_{i+1}} \frac{-1}{w^2}dw = \int_{t_i+1/w_i}^{t_{i+1}} d\tau$$

$$\Rightarrow \frac{1}{w_{i+1}} - \frac{1}{w_i} = t_{i+1} - \left(t_i + \frac{1}{w_i}\right) \Rightarrow \frac{1}{w_{i+1}} = I_{i+1} \tag{8}$$

On the next trial ($i$+1), $\varphi(\tau) = 1$ exactly when $\tau = t_{i+1}$, that is $\varphi_{i+1} = w_{i+1}I_{i+1} = 1$.

The complete learning rule can be summarized in a single equation:

$$dw = yw\frac{(1-\varphi)}{\varphi}d\tau - 1_{\varphi=1}w^2 d\tau \tag{9}$$

We remark that without loss of generality, a decreasing integrator (from 1 to 0) could be similarly constructed using the appropriate changes in bounds and sign in each rule.

### 2.3 Convergence proof

To simplify some of the following proofs, we will assume that there could be some cellular process marking the total change $\Delta w_i$ computed in (9) during interval $I_i$, but that after each event Y, only a small portion $\alpha$ of that change is actually applied to $w$ such that:

$$w_{i+1} = w_i + \alpha_i \Delta w_i \tag{10}$$

The effect of this assumption will be discussed at the end of the section.

**Lemma 1** Let $I_1, I_2, ..., I_n$ be a sequence of $n$ intervals, then given an appropriate learning rate schedule $\alpha_i = 1/i$, we will show that:

$$w_{n+1} = \frac{1}{n}\sum_{i=1}^{n} I_i^{-1} \tag{11}$$

**Proof:** If follows from section 2.2 that:

$$w_i + \Delta w_i = I_i^{-1} \tag{12}$$

After the first interval, since $\alpha_1 = 1$, the full correction is made, and therefore, by (12) and (10), we have that $w_2 = I_1^{-1}$ exactly. Now, let's assume (11) is true for $n$, then:

$$w_{n+2} = w_{n+1} + \frac{1}{n+1}\Delta w_{n+1} = \frac{nw_{n+1}}{n+1} + \frac{w_{n+1} + \Delta w_{n+1}}{n+1}$$

$$= \frac{1}{n+1}\sum_{i=1}^{n}I_i^{-1} + \frac{I_{n+1}^{-1}}{n+1} = \frac{1}{n+1}\sum_{i=1}^{n+1}I_i^{-1} \tag{13}$$

Therefore, (11) is true by induction.

**Theorem 1** Let $I_1, I_2, ..., I_n$ be a sequence of $n$ i.i.d. intervals with $E[I_i] = \mu$ and $Var(I_i) = \sigma$. Then given the learning rate schedule $\alpha_i = 1/i$, $w_n \to E[I^{-1}]$ in probability.

**Proof:** $\{w_n\}$ forms a sequence of means of the inverse of the delays $I_i$, $i \leq n$ and by the weak law of large number, $w_n \to E[I^{-1}]$ in probability. The model is learning the reciprocal of the harmonic mean of the intervals.

**Theorem 2** Let $I_1, I_2, ..., I_n$ be a sequence of $n$ i.i.d. intervals and let $0 < \alpha < 1$ constant. Then $\{1/w_n\}$ is an exponential moving harmonic average of the intervals.

**Proof:** Starting again from (10), substituting $\Delta w_i$ using (12), and using induction, we get:

$$w_{n+1} = w_n + \alpha\Delta w_n = w_n + \alpha(I_n^{-1} - w_n) = (1-\alpha)w_n + \alpha I_n^{-1} \tag{14}$$

Note that the use of (12) in this proof is not equivalent to placing the learning rate $\alpha$ directly into (9). While theorems would remain true for the first term (when the event occurs before $\varphi$ has saturated), the second term would push $\{w_n\}$ to converge to the harmonic mean of the event rate, i.e. the mean of the intervals:

$$\frac{1}{w_{i+1}} = (1-\alpha)\frac{1}{w_i} + \alpha I_{i+1} \tag{15}$$

While in general it could be preferable to learn the arithmetic mean of the interval, it is not clear which of the arithmetic or harmonic mean would best fit animal's data [13]. Some predictions related to the harmonic mean of the time intervals are made in section 3.3.

## 2.4 Noise and Weber's law for time

Adding Gaussian noise (Weiner process) to the accumulator and converting it into a stochastic differential system makes it a drift diffusion model (DDM, Figure 2, left panel) where $w$ is the rate of evidence under stimulus $x$ that the event Y will occur. Eq. (2) becomes

$$d\varphi = 1_{\varphi \neq 1} xw d\tau + cdW - y\varphi d\tau \tag{16}$$

where $cdW$ represent continuous time noise with distribution $N(0, c^2 dt)$. The decision time $DT$ (time of first occurrence of $\varphi > \theta$ for a threshold value $0 < \theta \leq 1$) has mean and variance

$$E[DT] = \frac{\theta}{w}\tanh\left(\frac{w\theta}{c^2}\right) \tag{17}$$

$$Var[DT] = \frac{\theta}{w}\frac{c^2}{w^2}g(y) \text{ where } y = -\frac{w\theta}{c^2} \tag{18}$$

as shown previously [14;15]. In the current system, we keep $\theta = 1$, $w = E[I^{-1}]$. If there is no noise ($c^2 \to 0$), then $tanh(w\theta/c^2) \to 1$. To maintain a linear temporal precision, that is to have $E[DT]/STD[DT] = z$, a constant independent of the distribution of $I_i$, one has to choose $c^2 = \beta^2 w$ for some constant $\beta > 0$. Thus $cdW = N(0, \beta^2 w d\tau)$ and we get:

$$E[DT] = \frac{\theta}{w} \tanh\left(\frac{\theta}{\beta^2}\right) \tag{19}$$

$$Var[DT] = \frac{\theta \beta^2}{w^2} g(y) \text{ where } y = -\frac{\theta}{\beta^2} \tag{20}$$

Since $w \approx E[T^1]$, we get that both the expected response time and the standard deviation on the expected response time are proportional to the estimated interval length encoded in $w$.

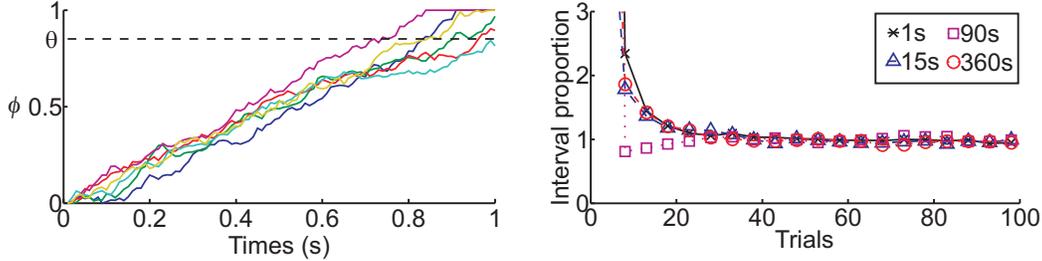

Figure 2: Results from fixed-interval temporal-conditioning *Experiment 1* ($\beta = .15$, $\theta = .85$, $\alpha = 0.1$) Left: Examples of few realizations of $\varphi$ between consecutive events for a unit trained on 1s fixed-interval. $\varphi$ is bounded below by 0 and the upper bound (1) is an absorbing process (see section 3). We say that there is a response when $\varphi > \theta$. Right: Estimated time encoded in the weight $w$ ($1/w$) as a function of trials for the first 100 trials. The number of trials to learn the delay is clearly independent of the time interval length.

### 2.5 Other properties

It is interesting to see that this model can have different and equally valid interpretations. First, at any moment, $\varphi/w$ is an estimate of the expected elapsed time since the last event and $(1-\varphi)/w$ is an estimate of the expected time until the next event, since $1/w$ is an estimate of the inter-event interval. Second, the threshold crossing of the integrator can be used to generate events under fixed weights, making it a generative model. Exact mean response time and timing precision also vary between animals in timing experiments. If one can determine a distribution for these variables, then $\beta$ and $\theta$ can be sampled from it and the model could technically be used to study both, the intra- and inter-subjects variability.

## 3 Results

To simplify the simulation, we considered the upper bound 1 on $\varphi$ as an absorbing process (eq. (16), once the accumulator is at 1, noise cannot bring it back down). We also put a lower bound of 0 on $\varphi$. This second change may affect (17) to (20), but the simulations below show that this change does not affect the time-scale invariance property of the model.

In temporal conditioning, reward is delivered regularly, independently of the animal behavior, and without any extra stimulus to mark the beginning of the interval. The reward delivery marks the beginning of a new interval. Under this simple protocol, animals learn that the reward comes regularly and also learn the approximate interval between rewards. They usually begin to actively try to get a reward after about 85% of the time interval (Figure 3, left panel) [1;16]. More complex conditioning procedures, interleaving random inter-trial intervals (without stimuli) and trial intervals (fixed temporal interval marked by a stimulus and terminating by a reward), can be used to evaluate the animal's timing precision. To do so, one can insert probe trials which are longer than trial intervals and which are not rewarded. Under such a procedure, one can see that the animal precision is always proportional to the interval length, with a precision of about 15% (Figure 3, right panel) [2;17]. One of the most important features a timing model should have is that the curves for different time scales should overlap when scaling the temporal axis appropriately (by the ratio of time scales). Finally, animals seem to learn the timing as fast as they acquire a response [1;2]. The following experiments shows that the present model (assuming $x$ is the stimulus marking the interval), can reliably reproduce behavioral data under these conditions, considering $\varphi > \theta$ as a response (Figure 2, left panel).

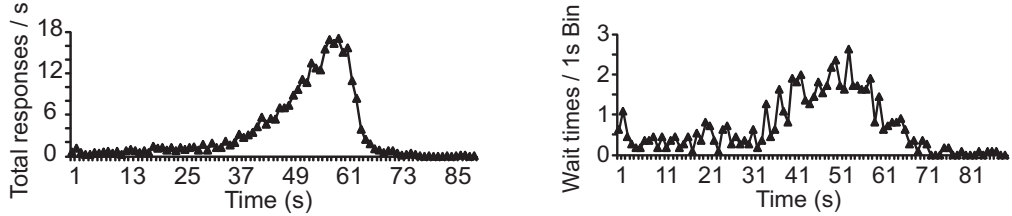

Figure 3: Unpublished empirical data coming from a rat trained on a 60s fixed-interval conditioning schedule. Left: Responses with respect to elapsed time. Right: Waiting time distribution before responding. The curves at different time scales have the same shape.

### 3.1 Experiment 1: Fixed-interval temporal-conditioning

In *Experiment 1*, we run the model through 200 consecutives trials of fixed-interval temporal-conditioning. We used noise with $\beta = 0.15$, since animal data show an average standard deviation over response time of about $\sigma = 0.15\mu$, and $\theta = 0.85$, since animals start to respond at about 85% of the time interval [1;2;16;17]. We also set $\alpha = 0.1$. We tested the model on four distinct intervals: 1s, 15s, 90s and 360s. In all four simulations, learning the timing took less than 20 trials (Figure 2, right panel). That is, learning the timing of the predictable event is easy [1;2] and independent of the time interval length to learn. After 100 trials, the timing encoded in the weight is within 3% error of the true interval length. Moreover, the response curve nicely reproduces the FI scallop (Figure 4, left panel) with the probability of mid-point response being at about 85% of the time interval, as well as its time-scale invariance.

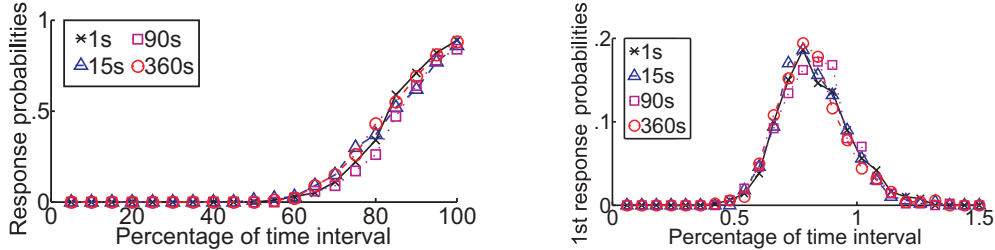

Figure 4: Model results from fixed-interval temporal-conditioning, scaled by the time interval length, comparable to animal data shown on Figure 3. In both cases, the model closely matches animal behavior. These results also show that the model is time-scale invariant. Left: Response probability with respect to elapsed time since last event, averaged over the last 100 trials of *Experiment 1*. In both experiments, the mid-response point is at about 85% of the interval, as set through the parameter $\theta$. Right: Probability distribution of the first response since last event from *Experiment 2*. The standard deviation over mean is about 15%, as set through the parameter $\beta$.

### 3.2 Experiment 2: Fixed-interval responses distributions

In *Experiment 2*, we ran the model through 500 probe trials using a fixed interval three times longer than the encoded interval. To do that, we directly encoded the desired delay into the weight $w$ and set the model's learning rate to $\alpha = 0$. We did that to show that the response probability distribution (Figure 4, left panel) and the first response time probability distribution (Figure 4, right panel) are time-scale invariant, in agreement with animal data [2;13]. The deviation in first response time is $\sigma = 0.15\mu$ and the mid-response point is at about 85% of the time interval, as set through $\beta$ and $\theta$.

### 3.3 Experiment 3: Normally distributed intervals

In *Experiment 3*, we run the model as in *Experiment 1* but using normally distributed intervals $N(a,(ab)^2)$, where $a \in \{1s, 15s, 90s\}$ and $b \in \{.01, .05, .1, .5\}$ instead of fixed intervals. When the environment variance is below the model variance ($b < \beta = .15$), the response curves are similar to those found using fixed intervals (Figure 4) and the mean encoded in the weight is close to the true mean of the interval (note that for $b < .15$, the true mean is close to the harmonic mean of the distribution). When the variance gets larger ($b = 0.5$), the model variance increases and the average encoded value converges to the harmonic mean of the interval distribution, hence the model tends to react more quickly than under fixed intervals. Animals seeing intervals based on a distribution made from the sum of a fixed and exponentially distributed intervals tend to underestimate the exponential part of the mean and react sooner [18], favoring the harmonic mean hypothesis.

# 4     Discussion

A complete literature review of interval timing psychological and neural models is outside the scope of this paper. But relations to some important models must be discussed.

## 4.1    Relation to SET and RET

The model presented here is based on a temporal accumulator like SET (scalar expectancy theory) [13], but without a central ticking clock (or pulse generator) nor any unbounded accumulator. It can be seen as a continuous, more neural, version of SET, requiring only a continuous stream of non-zero input ($x$) that can be provided by the presence of a stimulus or some steady-state working memory. Also in contrast to SET, the model adapts directly the accumulator rate to match the intervals it must learn, allowing it to work with bounded accumulators. Like SET, it is time-scale invariant, the accumulator remains more or less stable during gaps (when $x = 0$), and noise in the system is related to the intervals stored in memory. In contrast to SET, that samples an interval value only once from the memorized distribution, the present model continuously reads the interval length from the weight that encodes it by integrating $xw$ over time; a more neural form of memory.

The model is also related to RET (rate estimation theory) [13]. But RET is not an interval timing theory, it is only a conditioning theory and it also requires unbounded accumulators. It attempts to explain how stimulus, trial and intertrial intervals, and reinforcement interact and determine the acquisition of a conditioned response (but it does not tell *when*, within a trial, one would respond). The model presented here does not yet provide an explanation for these conditioning phenomena. Nevertheless, it is related to RET, since like RET, it depends on learning an estimate of events rate under different stimuli. The weight $w$ learned in the present model is therefore similar to the rate estimates λ in RET. By generalizing Eq. (16) to an evidence rates vector $w$ associated with a stimuli vector $x$, the model could serve as both an associative model of conditioning (what RET is) and a temporal map for timing predictions [2]. Since elapsed time and conditional expected time are always available within the unit's activity in our model, it is easy to think of a learning rule that could credit timing error to different stimuli based on such information. More work has yet to be done to account for the wide amount of empirical conditioning data RET is able to explain, but promising research in that direction are currently undertaken in our laboratory.

The present model not only subsumes the useful properties of SET, it does so in a more neurally realistic way than SET. Moreover, it allows us to think that it may eventually integrate all the great features of SET and RET within a unified adaptive decision model instead of two separate ones.

## 4.2    Relation to other build-up models

Two classes of neural temporal build-ups have been proposed: *(a)* a population of neurons whose individual activity increases (or decreases) as a function of elapsed time [7], and *(b)* a population of bistable neurons whose number of active neurons increases over time [6]. The present model is definitely in the first category.

General models of bistable neurons as in *(b)* were shown to have time scale-invariance, an important property to match behavioural data [6]. Almeida & Ledberg [6] model's has some commonality with the present model and even proposes some neural implementation. The main difference relies on the specific type of units used. In a population of bistable neurons, individual neurons contain no information about elapsed time; such information could only be inferred from the population read-out. In the present model, each individual unit has its own time estimate. This is a fundamental difference between representations *(a)* and *(b)*.

Finally, although Almeida & Ledberg [6] provide a learning rule, it is not clear how fast and easily it would converge, something clearly described and analyzed in our model.

Moreover, neural build-up of activity related to elapsed time as in *(a)* has been electrophysiologically observed in numerous brain regions [4;7;9-11]. Durstewitz [4] is making strong arguments in favour of this kind of model, but only few attempts at designing and evaluating such models appear in the literature. Reutimann and collaborators [7] developed such a model using spiking neurons, but the model presented here is more abstract, simpler, easier to analyze, has a clear and powerful learning rule, and has time-scale invariance.

The model presented here is an important milestone in our current understanding of interval timing by not only proposing an abstract neural model of interval timing that can accurately reproduce time scale invariance, but also by yielding activity that best matches with a majority of observed neural data. Most of all, it provides a simple, local learning rule that can be proved to learn timing rapidly and accurately.

### 4.3 Relationship to decision making

DDM parameters $w$, $\theta$, and $c$ are often fixed to fit response curves (reaction time) such as in two forced-choice tasks under very noisy inputs. In this paper, we show that the parameter $w$ can be learned from experience (by the subject) and could correspond to the mean reaction time under noise free observation of a stimulus ($x = 1$); $\theta$ is related to the fraction of the interval the animal waits before responding; and $c$ must be tightly linked to $w$. The animal reacts when it has accumulated enough evidence that the reward is about to occur (when $\varphi > \theta$) based on how long the conditioned stimulus is presented. Under a fixed interval schedule, the probability of reward to occur at any moment is related to the elapsed time.

## 5 Conclusion

The proposed model can learn any fixed time interval in $O(1/\alpha)$ trials (where $\alpha$ is the learning rate), independently of the interval length or of the previously learned value. With $\alpha = 0.1$, any time interval can be learned in about 20 trials. The timing error in the system (standard deviation) is proportional to the timed interval; the model is thus truly time-scale invariant. The response curves on fixed-interval temporal conditioning nicely fit the empirical data seen in animals. It does not only reproduce the population average response, but its whole distribution, allowing for analysis of individual realizations.

The model is extremely simple. The learning is based on a single weight and a single activity. No population of decays or eligibility traces, no delay-lines, nor any other form of large memory or history. The geometrical solution of the adaptive criterion makes it simple to understand and the model's property fit nicely SET theory, without the burden of unbounded accumulator and central clock. Moreover, the model's hyper-parameters $\alpha$, $\beta$, and $\theta$, can be estimated easily for any specific experimental preparation and can be used directly into the model without requiring tedious model fitting procedure. The results were not only showed true by simulations, but the properties were proved analytically.

An important aspect of this model is that the weights represent the contribution to event rates the connected stimulus is providing in accordance with RET theory. In agreement with SET theory, the noise is related to *reading* the memorized time value. The model makes three interesting predictions. First, if $w$ is related to the synaptic efficacies, then the variance of the noise in the synaptic transmission (noise in computing $xwd\tau+cdW$) must have variance proportional to the square root of the synaptic efficacies (since $c^2 = \beta^2 w$). Second, under normally distributed random intervals with standard deviation less than 15%, we should see little or no behavioral difference than under fixed-interval. Third, if the animals use a similar learning rule related to the harmonic mean of the intervals, than under higher variance, we should see an underestimate of the true mean of the distribution of the intervals.

Research to generalize the model to multiple stimuli while reproducing conditioning data accurately are undergoing in our lab. This paper presents a new simple way to use a drift-diffusion model for interval timing with a simple yet powerful learning rule. This may serve as a new basis to replace delay-lines representations of time and memory and eventually lead to learn powerful temporal associative maps.

## Acknowledgements

We are grateful to Elliot Ludvig for discussion in the development of the present work and for the empirical data used in Figure 3. We are also grateful to NSERC and Canada Research Chairs for their generous support.

## References

[1] Balsam, P. D., Drew, M. R., & Yang, C. (2002). Timing at the start of associative learning. Learning and Motivation 33(1): 141-155.

[2] Balsam, P. D. & Gallistel, C. R. (2009). Temporal maps and informativeness in associative learning. Trends Neurosci. 32(2): 73-78.

[3] Ivry, R. B. & Schlerf, J. E. (2008). Dedicated and intrinsic models of time perception. Trends Cogn Sci. 12(7): 273-280.

[4] Durstewitz, D. (2004). Neural representation of interval time. Neuroreport 15(5): 745-749.

[5] Buonomano, D. V. (2005). A learning rule for the emergence of stable dynamics and timing in recurrent networks. J.Neurophysiol. 94(4): 2275-2283.

[6] Almeida, R. & Ledberg, A. (2009). A biologically plausible model of time-scale invariant interval timing. J.Comput.Neurosci.

[7] Reutimann, J., Yakovlev, V., Fusi, S., & Senn, W. (2004). Climbing neuronal activity as an event-based cortical representation of time. J.Neurosci. 24(13): 3295-3303.

[8] Karmarkar, U. R. & Buonomano, D. V. (2007). Timing in the absence of clocks: encoding time in neural network states. Neuron 53(3): 427-438.

[9] Komura, Y., Tamura, R., Uwano, T., Nishijo, H., Kaga, K., & Ono, T. (2001). Retrospective and prospective coding for predicted reward in the sensory thalamus. Nature 412(6846): 546-549.

[10] Leon, M. I. & Shadlen, M. N. (2003). Representation of time by neurons in the posterior parietal cortex of the macaque. Neuron 38(2): 317-327.

[11] Lebedev, M. A., O'Doherty, J. E., & Nicolelis, M. A. (2008). Decoding of temporal intervals from cortical ensemble activity. J.Neurophysiol. 99(1): 166-186.

[12] Gold, J. I. & Shadlen, M. N. (2007). The neural basis of decision making. Annu.Rev.Neurosci. 30: 535-574.

[13] Gallistel, C. R. & Gibbon, J. (2000). Time, rate, and conditioning. Psychol.Rev. 107(2): 289-344.

[14] Wagenmakers, E. J., Grasman, R. P. P. P., & Molenaar, P. C. M. (2005). On the relation between the mean and the variance of a diffusion model response time distribution. Journal of Mathematical Psychology 49(3): 195-204.

[15] Bogacz, R., Brown, E., Moehlis, J., Holmes, P., & Cohen, J. D. (2006). The physics of optimal decision making: a formal analysis of models of performance in two-alternative forced-choice tasks. Psychol.Rev. 113(4): 700-765.

[16] Church, R. M. (2003). A Concise Introduction to Scalar Timing Theory. In W.H.Meck (Ed.), Functional and Neural Mechanisms of Interval Timing, pp. 3-22. Boca Raton: CRC Press.

[17] Gallistel, C. R. (2003). Time has come. Neuron 38(2): 149-150.

[18] Kirkpatrick, K. & Church, R. M. (2003). Tracking of the expected time to reinforcement in temporal conditioning procedures. Learn.Behav. 31(1): 3-21.